\documentstyle[preprint,prl,aps]{revtex}
\begin{document}
\draft
\title{Onsager Relations and Hydrodynamic Balance Equations
in 2D Quantum Wells}
\author{M. W. Wu$^1$, H. L. Cui$^1$, W. Sun$^1$, and S. Y. Wu$^2$}
\address{1. Department of Physics and Engineering Physics, Stevens
Institute of Technology, Hoboken, New Jersey 07030, USA}
\address{2. Department of Physics, University of Louisville, Louisville,
Kentucky 40292, USA}
\maketitle

\begin{abstract}

In this letter we clarify the role of heat flux in the hydrodynamic balance
equations in 2D quantum wells, facilitating the formulation of an Onsager
relation within the framework of this theory. We find that
the Onsager relation is satisfied within the
framework of the 2D hydrodynamic balance equation transport theory at
sufficiently high density. The condition of high density is consonant with
the requirement of strong electron-electron interactions for the validity of
our balance equation formulation.
\end{abstract}

\pacs{PACS number(s): 72.10.Bg, 72.20.Ht, 05.70.Ln}
\newpage

The Lei-Ting balance equation transport theory\cite{leiting,leihoring}
has achieved much success in hot-electron transport of homogeneous
semiconductors. This theory was subsequently generalized to deal with
weakly nonuniform, inhomogeneous systems\cite{lei}.
The resulting hydrodynamic balance equations
consist of continuity equation, momentum balance equation,
and energy balance equation. A salient feature of this hydrodynamic
approach is that it includes a
microscopic description of scattering in the form of
a frictional force function due to electron-impurity and electron-phonon
scattering, as well as an electron energy loss rate function due to
electron-phonon interaction. These hydrodynamic balance
equations have recently been applied to device
simulations\cite{cai1,cai2,cai3} and they have been further
developed to include phonon-drag effects\cite{wu1} and
applied to discussion of  thermoelectric power under both linear\cite{wu1}
and nonlinear\cite{wu2} transport conditions.

Until recently, the important issue concerning
the capability of this theory
to lead to the correct form of Onsager relations\cite{onsager}
and/or how to determine Onsager relations within the framework of this
theory has not been addressed.
There is even some misunderstanding that the energy flux
predicted by this theory is zero. It is well known that the Onsager
relation is a manifestation of microscopic irreversibility for any
statistical system near thermal equilibrium. Therefore any properly
formulated statistical physics model should satisfy this relation. It is
very easy to verify this relation in the framework of Kubo linear response
theory. Moreover, if one can determine the distribution function from
the Boltzmann equation, it is also straightforward to verify the Onsager
relation by calculating the pertinent moments of the distribution function.
However, for the traditional hydrodynamic
model,\cite{1,2,3,4,5,6,7,8,9,10,11,12} which is derived from Boltzmann
equation, verification has been elusive.
In fact, in a very recent article\cite{anile3}, it has been argued
that the Onsager relation breaks down in this model.

Recently we\cite{wu3}
clarified the role of heat flux in this theory, and, by introducing
the fourth balance equation, {\em ie.}, energy flux balance equation,
we were able to show
how to generate Onsager relations within the framework of this theory.
Moreover, we closely checked the Onsager relation predicted by this
theory for bulk semiconductors and found, that for any temperature,
when electron density is sufficiently high, the balance
equation theory satisfies Onsager relations exactly. To our knowledge,
this is the first set of hydrodynamic equations which has been
shown to obey Onsager relation exactly.

The purpose of the present letter is to clarify the role of heat flux in
hydrodynamic balance equations in quantum wells (and other 2D
systems), facilitating the
formulation of an Onsager relation within the framework of this
theory, and to verify the validity of this theory in
regard to the Onsager relation.

Following the procedure set forth in Ref.\ \cite{wu3},
the hydrodynamic balance
equations which describe a weekly inhomogeneous electron system
under the influence of an electron ${\bf E}$ and a small lattice
temperature gradient $\nabla T$ in two dimensional (2D) quantum
wells can be written as
\begin{eqnarray}
\label{eq1}
&&\frac{\partial n}{\partial t}+\nabla\cdot(n{\bf v})=0\;,\\
\label{eq2}
&&\frac{\partial}{\partial t}\langle {\bf J}\rangle+\nabla\cdot(
\langle {\bf J}\rangle{\bf v})
=-\frac{1}{m}\nabla u+\frac{{\bf f}}{m}\;,\\
\label{eq3}
&&\frac{\partial u}{\partial t}
+\nabla\cdot\langle{\bf J}_H\rangle=
{\bf v}\cdot\nabla u+\frac{1}{2}mv^2\nabla\cdot(n{\bf v})
+\frac{1}{2}mn{\bf v}\cdot\nabla v^2
-w-{\bf v}\cdot{\bf f}\;,
\end{eqnarray}
which describe continuity, momentum balance and energy balance
respectively. In these equations, the carrier drift velocity ${\bf v}$,
the electron temperature $T_e$, the average relative electron energy
$u$, the carrier density $n$ and the chemical potential $\mu$, together
with the particle flux $\langle{\bf J}\rangle$ and energy
flux $\langle {\bf J}_H\rangle$ are all field quantities
weakly dependent on the spatial coordinate, such that their spatial
gradients are small and therefore are retained only to first order
in Eqs.\ (\ref{eq1}-\ref{eq3}).  The energy
flux $\langle{\bf J}_H\rangle$
appearing in Eq.\ (\ref{eq3}) is just the energy flux predicted by
hydrodynamic balance equation theory. It relation to the other field
quantities in 2D case can be derived following
Ref.\ \cite{wu3} and be written as
\begin{equation}
\label{jhl}
\langle {\bf J}_H({\bf R})\rangle=2u({\bf R}){\bf v}({\bf R})
+\frac{1}{2}mn({\bf R})v^2({\bf R}){\bf v}({\bf R})\;.
\end{equation}
Substituting this relation, together with the relation of
density flux
\begin{equation}
\label{jav}
\langle {\bf J}({\bf R})\rangle=n({\bf R}){\bf v}({\bf R})\;,
\end{equation}
into Eqs.\ (\ref{eq2}) and (\ref{eq3}), one arrives at the original
hydrodynamic balance equations\cite{lei1}.

In order to obtain the Onsager relation in the framework of
balance equation theory, we need first derive the energy-flux
balance equation\cite{wu3}. This in the case
of 2D electrons in quantum wells, is given as
\begin{equation}
\label{eq4}
\frac{\partial}{\partial t}
\langle{\bf J}_H\rangle+\nabla\cdot\langle{\cal A}\rangle
=\langle{\bf B}\rangle+\frac{2}{m}eu{\bf E}+en{\bf E}\cdot{\bf v}{\bf v}
+\frac{1}{2}env^2{\bf E}+\frac{1}{2}v^2{\bf f}-w{\bf v}\;.
\end{equation}
The expression of $\langle{\bf B}\rangle$ is composed of two parts. One is
due to collisions with impurities ($\langle{\bf B}_i\rangle$), and the other
is due to interaction with phonons ($\langle{\bf B}_{ph}\rangle$). They
are given by
\begin{eqnarray}
\langle{\bf B}_i\rangle&=&2\pi n_i\sum_{\bf kq}|u({\bf q})|^2
|F({\bf q},z_0)|^2(\varepsilon_{
{\bf k}+{\bf q}}-\varepsilon_{\bf k})\frac{{\bf k}+{\bf q}/2}{m}\delta(
\varepsilon_{{\bf k}+{\bf q}}-\varepsilon_{\bf k}+{\bf q}\cdot{\bf v})
\nonumber\\
&&\hspace{1cm}\mbox{}\times\left[f(\frac{\varepsilon_{\bf k}-\mu}{T_e})-
f(\frac{\varepsilon_{{\bf k}+{\bf q}}-\mu}{T_e})\right]\nonumber\\
&&\mbox{}+2\pi n_i\sum_{\bf kq}|u({\bf q})|^2|F({\bf q},z_0)|^2
({\bf q}\cdot{\bf v}\frac{{\bf k}+{\bf q}}{m}+{\bf k}
\cdot{\bf v}\frac{{\bf q}}{m})\delta(\varepsilon
_{{\bf k}+{\bf q}}-\varepsilon_{\bf k}+{\bf q}\cdot{\bf v})\nonumber\\
&&\hspace{1cm}\mbox{}\times\left[f(\frac{\varepsilon_{\bf k}-\mu}{T_e})-
f(\frac{\varepsilon_{{\bf k}+{\bf q}}-\mu}{T_e})\right]\nonumber\\
&&\mbox{}+\pi n_i\sum_{\bf kq}|u({\bf q})|^2|F({\bf q},z_0)|^2
(\varepsilon_{{\bf k}+{\bf q}}
+\varepsilon_{\bf k})\frac{{\bf q}}{m}\delta(
\varepsilon_{{\bf k}+{\bf q}}-\varepsilon_{\bf k}+{\bf q}\cdot{\bf v})
\nonumber\\
&&\hspace{1cm}\mbox{}\times\left[f(\frac{\varepsilon_{\bf k}-\mu}{T_e})-
f(\frac{\varepsilon_{{\bf k}+{\bf q}}-\mu}{T_e})\right]\;,
\end{eqnarray}
and
\begin{eqnarray}
\langle{\bf B}_{ph}
\rangle&=&-4\pi\sum_{{\bf kQ}\lambda}|M({\bf Q},\lambda)|^2|I(iq_z)|^2
(\varepsilon_{{\bf k}+{\bf q}}-\varepsilon_{\bf k})
\frac{{\bf k}+{\bf q}/2}{m}\delta(\varepsilon_{{\bf k}+{\bf q}}-
\varepsilon_{\bf k}+\Omega_{{\bf Q}\lambda}-{\bf q}\cdot{\bf v})
\nonumber\\
&&\hspace{1cm}\mbox{}\times\left[f(\frac{\varepsilon_{\bf k}-\mu}{T_e})-
f(\frac{\varepsilon_{{\bf k}+{\bf q}}-\mu}{T_e})\right]\left[n(\frac{\Omega_
{{\bf Q}\lambda}}{T})-n(\frac{\Omega_{{\bf Q}\lambda}-{\bf q}\cdot{\bf v}}
{T_e})\right]\nonumber\\
&&\mbox{}-4\pi\sum_{{\bf kQ}\lambda}|M({\bf Q},\lambda)|^2|I(iq_z)|^2
({\bf q}\cdot{\bf v}\frac{{\bf k}+{\bf q}}{m}+{\bf k}\cdot{\bf v}
\frac{{\bf q}}{m})\delta(\varepsilon_{{\bf k}+{\bf q}}-
\varepsilon_{\bf k}+\Omega_{{\bf Q}\lambda}-{\bf q}\cdot{\bf v})
\nonumber\\
&&\hspace{1cm}\mbox{}\times\left[f(\frac{\varepsilon_{\bf k}-\mu}{T_e})-
f(\frac{\varepsilon_{{\bf k}+{\bf q}}-\mu}{T_e})\right]
\left[n(\frac{\Omega_{{\bf Q}\lambda}}{T})
-n(\frac{\Omega_{{\bf Q}\lambda}-{\bf q}\cdot{\bf v}}
{T_e})\right]\nonumber\\
&&\mbox{}-2\pi\sum_{{\bf kQ}\lambda}|M({\bf Q},\lambda)|^2|I(iq_z)|^2
(\varepsilon_{{\bf k}+{\bf q}}+\varepsilon_{\bf k})
\frac{{\bf q}}{m}\delta(\varepsilon_{{\bf k}+{\bf q}}-
\varepsilon_{\bf k}+\Omega_{{\bf Q}\lambda}-{\bf q}\cdot{\bf v})
\nonumber\\
&&\hspace{1cm}\mbox{}\times\left[f(\frac{\varepsilon_{\bf k}-\mu}{T_e})-
f(\frac{\varepsilon_{{\bf k}+{\bf q}}-\mu}{T_e})\right]\left[n(\frac{\Omega_
{{\bf Q}\lambda}}{T})-n(\frac{\Omega_{{\bf Q}\lambda}-{\bf q}\cdot{\bf v}}
{T_e})\right]\;.
\end{eqnarray}
The tensor ${\cal A}$ can be expressed as
\begin{equation}
\label{aa}
\langle{\cal A}\rangle=\frac{1}{2}[S({\bf R})+uv^2]{\cal I}+
3u{\bf v}{\bf v}+\frac{1}{2}mnv^2{\bf v}{\bf v}\;,
\end{equation}
with
\begin{equation}
\label{sr}
S({\bf R})=2\sum_{\bf k}\frac{k^4}{2m^3}f(\frac{\varepsilon_{\bf k}-\mu}
{T_e})\;.
\end{equation}
In these equations $\varepsilon_{\bf k}=k^2/2m$, and $f(x)=1/(e^x+1)$
represent the energy dispersion of 2D electrons and Fermi distribution
function separately, and ${\cal I}$ stands for the unit tensor.

The Onsager relation\cite{mahan} is concerned with the linear response of
the particle current $\langle{\bf J}\rangle$ and the heat
flux $\langle {\bf J}_Q\rangle$ near thermal equilibrium, which
flow as a result of generalized forces ${\bf X}_i$ on the system:
\begin{eqnarray}
\label{o1}
\langle{\bf J}\rangle&=&L^{11}{\bf X}_1+L^{12}{\bf X}_2\;,\\
\label{o2}
\langle{\bf J}_Q\rangle&=&L^{21}{\bf X}_1+L^{22}{\bf X}_2\;,
\end{eqnarray}
with ${\bf X}_1=-\frac{1}{T}\nabla (\mu+e\phi)$ and ${\bf X}_2=\nabla(1/T)$.
The Onsager relation states that
\begin{equation}
L^{12}=L^{21}\;.
\end{equation}
The heat flux $\langle{\bf J}_Q\rangle$ relates to the energy flux in
Eq.\ (\ref{jhl}) through
\begin{equation}
\langle{\bf J}_Q\rangle=\langle{\bf J}_H\rangle-\mu\langle{\bf J}\rangle\;.
\end{equation}
The fluxes $\langle{\bf J}\rangle$ and $\langle{\bf J}_H\rangle$ have already
been defined by Eqs.\ (\ref{jav}) and (\ref{jhl}).
Our task is to express them in terms of linear response in the form of
Eqs.\ (\ref{o1}) and (\ref{o2}). To this end, we consider
electron transport in a quantum well which is
grown along $z$-direction), in the presence of
a small lattice temperature gradient and a small electric field along the
$x$ direction: $\nabla T=(\nabla_xT,0,0)$ and ${\bf E}=(E_x,0,0)$
respectively. Therefore $T_e=T$, and ${\bf v}=(v_x,0,0)$ is also small.
The system is near equilibrium. Following the same steps as in
Ref.\ \cite{wu3}, we can derive the first relation (\ref{o1})
from the momentum balance equation (\ref{eq2}) and the second relation
(\ref{o2}) from the energy-flux balance equation (\ref{eq4}), by treating
Eqs.\ (\ref{eq2}) and (\ref{eq4}) to first order in the small quantities
$E_x$, $\nabla_x T$ and $v_x$, and have
\begin{eqnarray}
L^{11}&=&\frac{T}{\rho e^2}\;,\\
\label{l12}
L^{12}&=&\frac{T^2}{\rho e^2}\left[2\frac{F_1(\zeta)}
{F_0(\zeta)}-\zeta\right]\;,\\
\label{l21}
L^{21}&=&\frac{T^2}{\rho e^2}\left[-\frac{\tau\rho e^2}{m}2
\frac{F_1(\zeta)}{F_0(\zeta)}-\zeta\right]\;,\\
L^{22}&=&-\frac{\tau T^3}{m}\left[3\frac{F_2(\zeta)}
{F_0(\zeta)}-2\zeta\frac{F_1(\zeta)}{F_0
(\zeta)}\right]-\frac{\zeta T^3}{\rho e^2}\left[2
\frac{F_1(\zeta)}{F_0(\zeta)}-\zeta\right]\;.
\end{eqnarray}
In deriving these equations, we have used the relations
\begin{equation}
\label{n}
n=2\sum_{\bf k}f[(\varepsilon_{\bf k}-\mu)/T_e]=\frac{2m}{\pi}T
F_0(\zeta)\;,
\end{equation}
and
\begin{equation}
\label{u}
u=2\sum_{\bf k}\varepsilon_{\bf k}f[(\varepsilon_{\bf k}-\mu)/T_e]
=\frac{2m}{\pi}T^2F_1(\zeta)\;,
\end{equation}
with $\zeta\equiv\mu/T$ and $F_\nu(y)=\int_0^\infty x^\nu[\exp(x-y)
+1]^{-1}dx$. Here,
\begin{equation}
\rho=-\frac{f_x}{n^2e^2v_x}=-\frac{f_x}{ne^2\langle J_x\rangle}
\end{equation}
is the resistivity and independent of $v_x$ ($\langle J_x\rangle$).
The expression of it can be found in Ref.\ \cite{leihoring}. Further,
\begin{equation}
1/\tau\equiv\frac{\langle B_x\rangle}{n({\bf R})\langle J_H^x\rangle}
\end{equation}
is also independent of $v_x$ (therefore $\langle J_H^x\rangle$). Comparing
Eq.\ (\ref{l12}) with Eq.\ (\ref{l21}), we find that the condition
under which the Onsager relation holds is given by
\begin{equation}
\label{i}
I\equiv-\frac{\tau\rho e^2}{m}=1
\end{equation}

We have closely examined Eq.\ (\ref{i}) for a  GaAs-based
(10\ nm well width along the $z$-direction) quantum well structure.
(We have also considered other quantum wells of different well width
and found that the tendency of $I$ approaching to ``1''
is not very sensitive on the width of quantum well.)
We only take the lowest subband occupation into account.
Both $\rho$ and $\langle B_x\rangle$ are composed of contributions due to
electron-impurity, electron--LO-phonon, and
electron--acoustic-phonon scatterings
(with the electron--acoustic-phonon scatterings due to longitudinal mode
acoustic phonons via deformation potential and piezoelectric
interactions, and transverse mode via piezoelectric interaction).
We have examined each scattering contribution in detail to check
Eq.\ (\ref{i}) separately for each interaction. It is clear that if
$I_i\equiv-\frac{e^2\rho_i/m}{(1/\tau)_i}=1$ is satisfied for
each interaction, we have $-\frac{e\sum_i\rho_i/m}{\sum_i(1/\tau)_i}=1$.
Moreover, this procedure is advantageous in that each $I_i$ is
independent of impurity concentration and parameters of the electron-phonon
interaction matrixes.

The expressions for $I$ obtained from the balance equations are given by
\begin{equation}
\label{iei}
I_{ei}=\frac{\sum_{\bf q}q^2|u({\bf q})|^2|F({\bf q},z_0)|^2
[\frac{\partial}{\partial \omega}
\Pi_2^\varepsilon({\bf q},\omega)]|_{\omega=0}}{2(\frac{u}{n})
\sum_{\bf q}q^2|u({\bf q})|^2|F({\bf q},z_0)|^2
[\frac{\partial}{\partial \omega}
\Pi_2({\bf q},\omega)]|_{\omega=0}}\;,
\end{equation}
due to electron-impurity scattering; and
\begin{eqnarray}
I_{e-ph}(\lambda)&=&\frac{\sum_{\bf Q}|M({\bf q},\lambda)|^2|I(iq_z)|^2
\Omega_{{\bf Q}\lambda}(\varepsilon_{\bf q}+\Omega_{{\bf Q}
\lambda})n^\prime(\frac{\Omega_
{{\bf Q}\lambda}}{T})\Pi_2({\bf q},\Omega_{{\bf Q}\lambda})}{2
(\frac{u}{n})\sum_{\bf q}|M({\bf Q},\lambda)|^2|I(iq_z)|^2
\frac{{\bf q}^2}{m}n^\prime(\frac{\Omega_{{\bf Q}\lambda}}{T})
\Pi_2({\bf q},\Omega_{{\bf Q}\lambda})}\nonumber\\
\label{ieph}
&&\mbox{}+\frac{-\sum_{\bf Q}|M({\bf Q},\lambda)|^2|I(iq_z)|^2
\frac{{\bf q}^2}{m}
n^\prime(\frac{\Omega_{{\bf Q}\lambda}}{T})\Pi_2^\varepsilon({\bf q},
-\Omega_{{\bf Q}\lambda})}{2
(\frac{u}{n})\sum_{\bf Q}|M({\bf Q},\lambda)|^2|I(iq_z)|^2
\frac{{\bf q}^2}{m}n^\prime(\frac{\Omega_{{\bf Q}\lambda}}{T})
\Pi_2({\bf q},\Omega_{{\bf Q}\lambda})}\;,
\end{eqnarray}
due to electron-phonon scattering, for phonons of mode $\lambda$.
$I_{e-ph}(\lambda)$ is further composed of contributions due to
electron--LO-phonon scattering, $I_{e-LO}$; due to  electron--longitudinal
acoustic phonons by deformation potential coupling, $I_{edl}$;
and by piezoelectric interaction, $I_{epl}$; and due to
electron--transverse acoustic phonons by piezoelectric
interaction, $I_{ept}$. In these equations, $\Omega_{{\bf Q}\lambda}$
is the phonon frequency of wave vector ${\bf Q}\equiv({\bf q},q_z)
\equiv(q_x,q_y,q_z)$ and mode $\lambda$; $u({\bf q})$, the electron-impurity
interaction potential; and $M({\bf q},\lambda)$, the electron-phonon
interaction matrix element. $F({\bf q},z_0)$ and $I(iq_z)$ are form
factors of electron-impurity and electron-phonon interaction
respectively, with $z_0$ standing for the position of impurity.
$n(x)=(e^x-1)^{-1}$ stands for the Bose distribution.
$\Pi_2({\bf q},\lambda)$ denotes the imaginary
part of electron density-density correction function.
$\Pi_2^\varepsilon$ is defined by
\begin{equation}
\Pi_2^\varepsilon({\bf q},\omega)=2\pi\sum_{\bf k}\varepsilon_{\bf k}
\delta(\varepsilon_{{\bf k}+{\bf q}}-\varepsilon_{\bf k}+\omega)
\left[f\left(\frac{\varepsilon_{\bf k}-\mu}{T}\right)-
f\left(\frac{\varepsilon_{{\bf k}+{\bf q}}-\mu}{T}\right)\right]\;.
\end{equation}
For the LO phonon, $\Omega_{{\bf Q},LO}=\Omega_0=35.4$\ meV, and the
Fr\"ohlich matrix element is $|M({\bf Q},LO)|^2=e^2(\kappa_\infty^{-1}
-\kappa^{-1})\Omega_0/(2\varepsilon_0Q^2)\propto 1/Q^2$. (Since the
constants in the matrix elements cancel in Eq.\ (\ref{ieph}),
in the following we only specify their relation to $Q$.) The matrix
element due to longitudinal deformation potential coupling is
$|M({\bf Q},dl)|^2\propto Q$, that due to longitudinal
piezoelectric interaction is $|M({\bf Q},pl)|^2\propto(q_xq_yq_z)^2/Q^7$,
and for the two branches of independent transverse piezoelectric
interaction: $\sum_{j=1,2}|M({\bf Q},pt_j)|^2\propto (q_x^2q_y^2+q_y^2q_z^2
+q_z^2q_x^2-(3q_xq_yq_z)^2/Q^2)/Q^5$. For acoustic phonons $\Omega_{{\bf Q}
\lambda}$ can be written as $v_{s}Q$, with the longitudinal sound speed
$v_s$ being 5.29$\times 10^3$\ m/s, and the transverse sound speed being
2.48$\times 10^3$\ m/s. The effective mass of electron is $0.07m_e$,
with $m_e$ denoting the free electron mass.

We present the results of  our numerical calculations in Fig.\ 1 to
Fig.\ 4, where contributions to  $I$ due to the various interactions
discussed above are plotted against electron density for several different
temperatures (the results of $I_{epl}$ are similar to those of
$I_{ept}$, so we only plot one as representative).
As it is generally believed that the contribution of acoustic
phonons is important only at low temperature, while the contribution
of LO phonons is dominant at high temperature, our temperatures are chosen
as 10, 20, and 40\ K for the former, and 50, 300, 500, and 1000\ K for
the letter. Impurity scattering is present at all temperature,
so we take $T=$10, 50, 100, 300, and 1000\ K in Fig.\ 1. From these
figures it is evident that, for any temperature, when electron density is
sufficiently high $I$ is exactly unity, indicating that the Onsager relation
holds. This is consistent with the understanding that the Lei-Ting
balance equation theory holds only for strong electron-electron
interaction\cite{chen1,chen2}.

\acknowledgements

This research is
supported by the U.S. Office of Naval Research
(Contract No. N66001-95-M-3472),
the U.S. Army Research Office (Contract NO. DAAH04-94-G-0413), and
partially by the National Science Foundation (Grant No. OSR-9452895).

\begin{figure}
\caption{$I$ due to electron-impurity scattering is plotted as a function of
electron density for several different temperatures}
\end{figure}

\begin{figure}
\caption{$I$ due to electron--LO-phonon scattering is plotted
as a function of electron density for several different temperatures}
\end{figure}

\begin{figure}
\caption{$I$ due to electron--longitudinal acoustic-phonon scattering via
deformation potential coupling is plotted
as a function of electron density for several different temperatures}
\end{figure}

\begin{figure}
\caption{$I$ due to electron--transverse acoustic-phonon scattering via
piezoelectric interaction is plotted
as a function of electron density for several different temperatures}
\end{figure}

\end{document}